# Ground-Based Radar Tracking of Near-Earth Objects With VLBI Radio Telescopes: 2024 MK Test Case

*Oliver White, Guifré Molera Calvés, Shinji Horiuchi, Ed Kruzins,*
*Edwin Peters, and Nick Stacy*

*Abstract* – The Southern Hemisphere Asteroid Research Project is an active and informal entity comprising the University of New South Wales, the University of Tasmania, the University of Western Australia, and the Curtin University, which performs asteroid research in collaboration with federal agencies, including the Commonwealth Scientific and Industrial Research Organisation and the National Aeronautics and Space Administration (JPL). Since 2015, we have used the Australian infrastructure to characterize more than 50 near-Earth asteroids through bistatic radar observations. On 29 June 2024, we used four very long baseline interferometer (VLBI) radio telescopes to follow the close approach of 2024 MK to the Earth. In this paper, we describe the detections and the analysis of VLBI and how these observations can help to improve the understanding of its composition and orbit characterization.

## 1. Introduction

Radar observations have been conducted from the northern hemisphere for decades in the study of Solar System objects [1, 2], including the refinement of planetary ephemerides to support interplanetary spacecraft missions and the detection of planets, comets, asteroids, moons, and Saturn's rings. These observations enabled numerous developments in planetary science, such as the determination of topographical features, orbital motion, and surface composition of these objects.

Radar observations of near-Earth asteroids (NEAs) have been conducted in the Northern Hemisphere since 1968 when Icarus (1566) was detected [3]. The benefits of these observations include recovering and refining known NEA orbits, which are up to multiple orders of magnitude greater than optical observations alone [1]. This orbit determination can improve spacecraft navigation when undertaking missions to these objects. Radar observations of NEAs also have planetary defense applications in the context of better predicting close approaches to Earth [4].

The echoes received from bistatic radar observations of NEAs can also yield insights into their surface and rotational properties, as well as whether they are binary objects. In 2018, the close approach of 2003 SD220 was used by Horiuchi et al. [5] to detect radar echoes of the NEA with the Australia Telescope Compact Array (ATCA). From this, an elongation of 1:8 was suggested, consistent with light curve data implying a rotation period of $285 \pm 5$ hours [6].

Before 2015, planetary radar observations were conducted primarily in the Northern Hemisphere. The Goldstone 70 m Deep Space Station 14 (DSS-14), the 34-m DSS-13 antennas, Arecibo, and the Green Bank Telescope were used extensively, with DSS-14 and Arecibo, respectively, capable of detecting 131 and 253 NEAs in 2015 when operating in monostatic mode [7]. Giorgini et al. [8] and Naidu et al. [7] described in detail the potential benefits of implementing radar observations of NEAs in various locations, particularly in the Southern Hemisphere.

To this end, the first observations from what would become the Southern Hemisphere Asteroid Research Program (SHARP) were conducted in 2015 [9]. The Canberra Deep Space Communication Complex antennas were used for transmission, with the Parkes Radio Telescope and ATCA acting as receivers. Many NEAs have been observed from the Southern Hemisphere in the years following the implementation of this setup [5, 9–13], demonstrating the Southern Hemisphere's significant contribution to the study of these objects.

Additionally, the University of Tasmania (UTAS) joined SHARP in March 2021, contributing with their continental-scale array of VLBI-equipped radio telescopes, as described in White et al. [14]. Figure 1 details the radio telescopes used for this purpose: the 12 m AuScope radio telescopes at Hobart (Hb), Katherine (Ke), and Yarragadee (Yg), as well as the 26 m at Hobart (Ho) and 30 m at Ceduna (Cd).

In this paper, we describe the use of the Cd, Ke, Yg, and Hb antennas to follow the close approach of 2024 MK to the Earth, with multistatic radar observations conducted on June 29, 2024. The analysis of the detections with VLBI details how these observations can aid in determining the composition and orbit of 2024 MK.

Oliver White is with University of Tasmania, 1 Churchill Avenue, Sandy Bay, Tasmania 7005, Australia; e-mail: oliver.white@utas.edu.au.

Guifré Molera Calvés is with University of Tasmania, 1 Churchill Avenue, Sandy Bay, Tasmania 7005, Australia; e-mail: guifre.moleracalves@utas.edu.au.

Shinji Horiuchi is with Commonwealth Scientific and Industrial Research Organisation/NASA Canberra Deep Space Communication Complex, Tidbinbilla, ACT 2620, Australia; e-mail: shinji.horiuchi@csiro.au.

Ed Kruzins is with University of New South Wales Canberra/Commonwealth Scientific and Industrial Research Organisation, Canberra 2610, Australia; e-mail: e.kruzins@unsw.edu.au.

Edwin Peters is with University of New South Wales Canberra, Canberra 2610, Australia; e-mail: edwin.peters@unsw.edu.au.

Nick Stacy is an independent researcher, Adelaide, Australia; e-mail: nick@bluemoonlab.au.

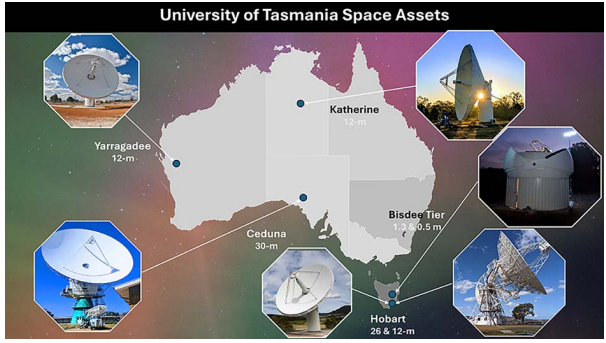

Figure 1. Location of the University of Tasmania antennas used for multistatic tracking of near-Earth objects.

## 2. Methodology

The UTAS radio telescopes and ATCA acted as multistatic receivers, with the DSSs at the Canberra Deep Space Communication Complex in Tidbinbilla transmitting a 20-kW continuous radio signal. The DSSs transmitted in the C-band at 7159.4 MHz, using left-hand circularly polarized electromagnetic continuous waves. Both Ho and Cd featured circularly polarized receivers at the C-band, while Hb, Ke, and Yg used linearly polarized receivers. As part of the VLBI community, all UTAS radio telescopes are equipped with standard VLBI backends, data acquisition systems, and hydrogen masers as clock standards. A digital baseband converter, consisting of three units, was used to sample analogue signals at each station, recording at 8 bits per sample, with the data recorded into standard data acquisition systems for further data processing [15]. This recording system is used in VLBI observations. Additionally, Hb, Ke, and Cd feature software-defined radio units, which can record up to 32 bits per sample.

The transmitted frequency was different at each receiver due to the Doppler shift. To a margin of error of tens of hertz, this was approximately related to the radial velocity between the NEA and the receiving station by,

$$f_D = \frac{(v_{tx-a} + v_{a-rx})f_{tx}}{c_0}, \quad (1)$$

where $f_{tx}$ and $f_D$ were the transmission and Doppler frequency shift, respectively, $v_{tx-a}$ was the radial velocity between the transmitter and the NEO, $v_{a-rx}$ was the radial velocity between the NEO and receiver, and $c_0$ was the speed of light. The signal transmitted from the transmitting station was already pre-Doppler compensated so that the arriving frequency at Cd was static over time. The frequency of arrival at the other stations depends on the relative motion between them, primarily caused by the Earth's rotation. The Doppler shift prediction at each site relative to Cd is shown in Figure 2.

We used the Spacecraft Doppler tracking (SDtracker) software version 2.0.14 [15] for data processing. The data processing pipeline for these observations is discussed in detail in White et al. [14]. In summary, SDtracker took the raw data recorded at each station and compensated for any

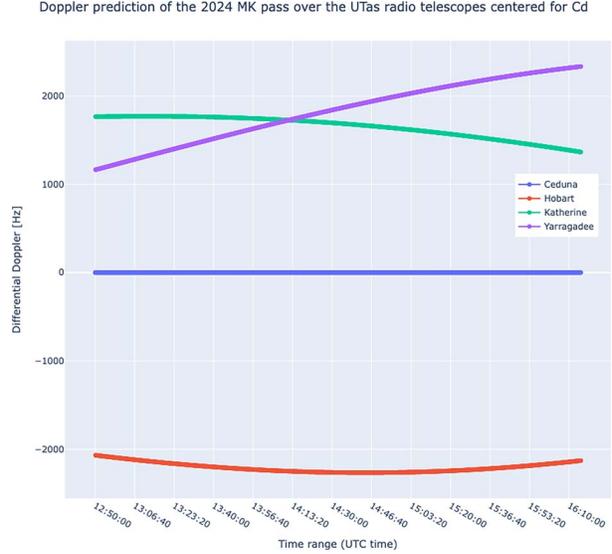

Figure 2. Prediction of the Doppler shift of the radar echo at the UTAS radio telescopes during the June 29, 2024, observation of 2024 MK. As the transmitted signal was already pre-Doppler compensated for Ceduna, this shows the Doppler shifts at the other stations.

Doppler shift arising from the relative motions of the NEA and the telescope. The final spectrum was created by integrating both coherent and noncoherent components.

The transmission frequency was shifted in time to receive radar echoes at Cd centered at 7159.45 MHz. SDtracker enabled the echoes to be centered for each station by using a phase polynomial, which calculates the expected received frequency due to the Doppler shift. The Python Astroquery module was used to retrieve values from the JPL Horizons online Solar System data service [17], which enabled the calculation of the Doppler shift using (1) with a radial velocity approximation. This was used to calculate the polynomial in frequency and phase for use in SDtracker. We were also supplied with predict files that contained each station's expected received frequency, which was alternatively used to generate these polynomials. This was based on a high-precision relativistic Doppler correction, created by the Jet Propulsion Laboratory's "On-Site Orbit Determination" tool, that is accurate to the second order [16].

The output from SDtracker enabled the creation of various plots, including the summed frequency spectrum centered on the expected radar echo and waterfall plots of frequency and time. The signal strengths were normalized by the noise floor and scaled by the standard deviation of the noise, giving a sigma value of the peak signal-to-noise ratio (SNR).

Using the radar range equation for a bistatic radar system [1], the SNR was given by,

$$SNR = \frac{P_t G_t G_r \lambda^2 \sigma (\Delta t)^{1/2}}{(4\pi)^3 R^4 k T_{sys} (\Delta f)^{1/2}} \quad (2)$$

where $P_t$ was the transmitted power, $G_t$ and $G_r$ were the respective gains, $\lambda$ was the transmitted wavelength, $\sigma$ was the radar cross section of the NEA and $\Delta t$ was the total integration time. $R$ was the range of the NEA from the ground stations, $T_{sys}$ was the receiving antenna's system temperature, and $\Delta f$ was the frequency resolution. If other terms are known, this would enable the calculation of the radar cross section of 2024 MK.

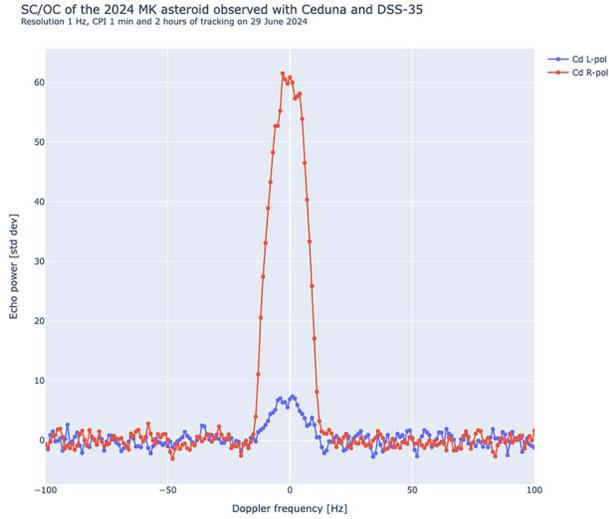

Figure 3. The sum of spectra in left (opposite) and right (same) circular polarizations from the radar observation of 2024 MK from Cd on June 29, 2024, with a frequency resolution of 1 Hz used, over a total scan length of 136 minutes.

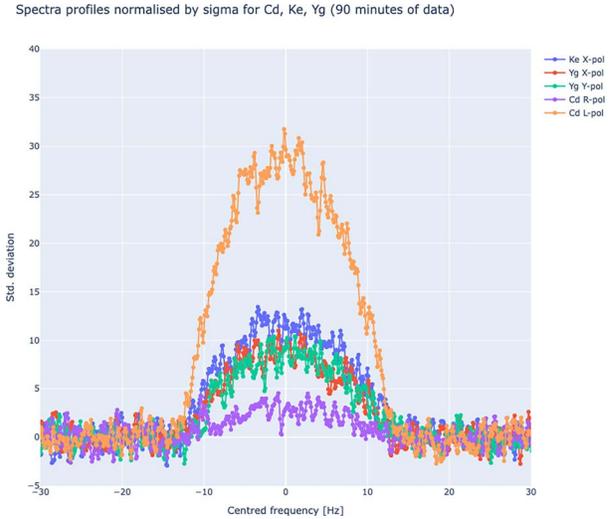

Figure 4. Ninety-minute sum of spectra in all polarizations recorded from the radar observation of 2024 MK from Ke, Yg, and Cd on June 29, 2024.

Additionally, following the successful centering of the signal using SDtracker, the broadening of the radar echo was measured. Considering only the impacts of the asteroid's properties, such as its rotation and surface, the Doppler broadening of the radar echo was given by,

$$B = \frac{4\pi D(\phi)\cos(\delta)}{\lambda P},\qquad(3)$$

where $P$ was the rotation period, $D(\phi)$ was the diameter at rotation phase $\phi$, and $\delta$ was the subradar latitude. This Doppler broadening arises from the NEA's relative rotation about its center of mass, with the leading and trailing edges having a greater magnitude of broadening. Depending on the subradar latitude, this enabled the estimation of the rotation period of the NEA.

## 3. Results

On June 16, 2024, ATLAS-Sutherland, located in South Africa and funded by the National Aeronautics and Space Administration, discovered 2024 MK, which is an NEA. It was predicted to make its closest approach to Earth on June 29, 2024, within 0.77 lunar distances, and its diameter was estimated to be between 120 and 260 m.

SHARP observations of 2024 MK were conducted on June 26, 28, and 29. The first two were during another DSS-43 and ATCA observation time scheduled for 2011 UL21, with the June 29 session a Target of Opportunity DSS-35 observation for 2024 MK. This final experiment included four of the UTAS radio telescopes as receivers: Cd, Ke, Yg, and Hb, as well as ATCA.

The antennas recorded from 11:00 UTC to 16:00 UTC, with Cd and ATCA recording both circular polarizations and the 12-m antennas recording both linear polarizations. The transmitted wave was continuous, except for breaks due to overhead aircraft, centered for reception at Cd at 7159.45 MHz. This signal was transmitted as a left-hand circularly polarized wave with a total power of 20 kW.

Aside from recording issues at Hb, the radar echo was definitively detected at Cd, Ke, Yg, and ATCA. Cd recorded a SNR of 62-sigma in the opposite (left) circular polarization and 8-sigma in the same (right) circular polarization, detailed in Figure 3. Figure 4 shows the signals detected at Cd, Ke, and Yg.

The Doppler compensation was performed accurately at each station, as can be seen in the negligible drift in Figure 5. This is despite the significant predicted Doppler shift throughout the observation, as demonstrated in Figure 2.

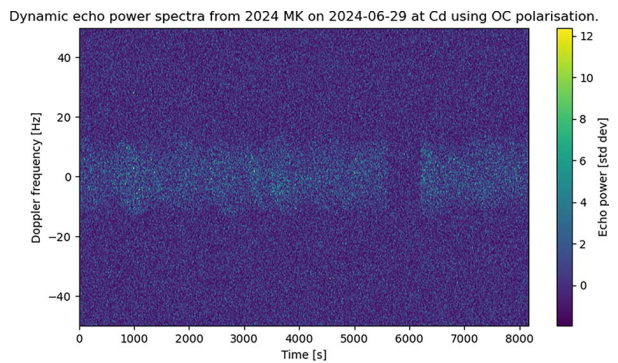

Figure 5. Waterfall spectra plot from the radar observation of 2024 MK from Cd on June 29, 2024, using the left (opposite) circular polarization [14].

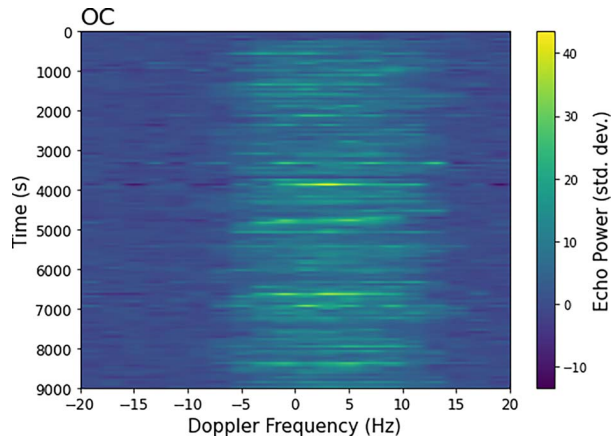

Figure 6. Waterfall spectra plot from the radar observation of 2024 MK from ATCA on June 29, 2024, using the left (opposite) circular polarization.

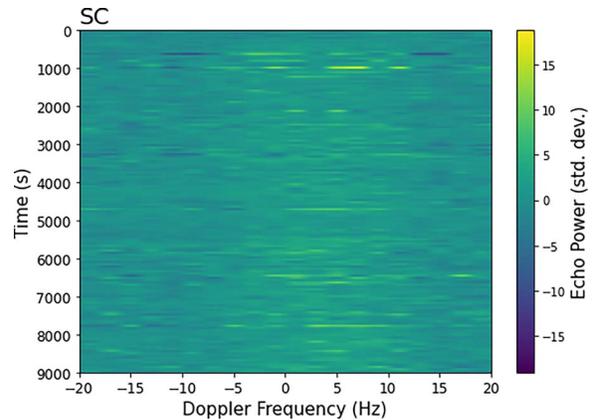

Figure 7. Waterfall spectra plot from the radar observation of 2024 MK from ATCA on June 29, 2024, using the right (same) circular polarization.

The expected Doppler broadening of the signal cannot be calculated as there are no known values of its rotation period. However, using (3) and assuming the NEA is observed edge-on so that the subradar latitude is 0, then the observed Doppler broadening is consistent with a period between approximately 1400 and 3100 s, based on a diameter estimate of 120 to 260 m. A nonzero subradar latitude would result in a smaller rotation period.

The SNRs of the different polarizations at Cd suggest a circular polarisation ratio (SC/OC) of 0.12. As described in Benner et al. [18] the SC/OC values are a measure of the roughness of the asteroid's surface and are correlated with its spectral class. Spectroscopic analysis has determined that 2024 MK is an S-class asteroid [19]. We do not have an estimate of the uncertainty of our SC/OC value; however, the analysis by Benner et al. [18] suggests that S-class NEAs have an SC/OC ratio of 0.270 and a standard deviation of 0.079. Thus, our value is consistent within $2\sigma$.

Figures 6 and 7 are ATCA waterfall plots of this observation, which indicate the rotation pattern of 2024 MK. Consistent with Cd, the signal is weaker in the right circular polarization than in the left circular polarization. Qualitatively, the rotation period this suggests would be consistent with the prediction based on the Doppler broadening.

## 4. Conclusion

As part of SHARP, we have demonstrated the ground-based radar tracking of 2024 MK using VLBI radio telescopes and the ability to accurately calculate and compensate for the Doppler shift at each station. This has enabled measurements of the radar echo and estimation of the NEA's physical parameters, including its rotation period.

Further analysis by cross-correlating the signals received at each station will enable the determination of the differential times of arrival. Not only will this improve the orbit determination through more precise timing of the Doppler shift, and thus radial velocity, but also the shift of the asteroid in time, enabling the derivation of the state vector components. This VLBI analysis will require the use of near-field delay models for 2024 MK, as well as consideration of ionospheric and tropospheric corrections. An existing delay model and processing methodology have been developed for spacecraft VLBI observations [20], which is currently being adapted for NEAs.

Additionally, we aim to use our results to perform tomographic imaging of the NEA, further characterizing the properties of 2024 MK.